\documentclass{PoS}
\usepackage[utf8]{inputenc}

\title{Astroparticle Physics in Hyper-Kamiokande}

\ShortTitle{Astroparticle Physics in Hyper-Kamiokande}

\author{\speaker{Jost Migenda}%
       \thanks{for the Hyper-Kamiokande Proto-Collaboration}\\
      University of Sheffield\\
      E-mail: \email{jmigenda1@sheffield.ac.uk}}

\abstract{
Hyper-Kamiokande (Hyper-K) is a proposed next-generation general purpose neutrino detection experiment.
It comprises an underground water Cherenkov detector that will be more than 8 times as large as the highly successful Super-Kamiokande and use significantly improved photodetectors with the same 40\,\% photocoverage.
The resulting sensitivity improvements will particularly benefit astroparticle physics at low energies.\\
This talk will give an overview over Hyper-K and present its projected physics reach in the areas of supernova neutrinos, solar neutrinos and indirect dark matter searches, based on the current design report. It will also discuss additional sensitivity improvements if the second detector is built in Korea in a location with a higher overburden.}

\FullConference{The European Physical Society Conference on High Energy Physics\\
		 5-12 July\\
		 Venice, Italy}

\begin{document}

\section{The Hyper-Kamiokande Experiment}
Hyper-Kamiokande (HK, \cite{PublicDR}) is a proposed next-generation general purpose neutrino detection experiment whose broad physics programme covers many areas of particle and astroparticle physics. Based on the proven technology of (Super-)Kamiokande, its much higher detector volume and additional improvements in key areas like photosensors and near/intermediate detectors make HK a straightforward yet powerful extension of the very successful (Nobel Prizes~2002~and~2015) Japan-based neutrino programme.

These proceedings will give an overview over the experiment in the remainder of this section and will then discuss expected sensitivities in the main areas of astroparticle physics --~supernova neutrinos (chapter~\ref{ch:sn}), solar neutrinos (chapter~\ref{ch:solar}) and indirect dark matter searches (chapter~\ref{ch:dm})~-- based on the design report published in 2016~\cite{PublicDR}.

\subsection{Detector Design}

HK consists of an underground water Cherenkov detector that will be located about 8\,km south of Super-Kamiokande (SK) in the Tochibora mine with an overburden of 1750\,m.w.e.
The detector will be cylindrical (60\,m high and 74\,m in diameter) and have a fiducial (total) mass of 187 (260) kton, making it more than 8 (5) times as large as its predecessor.
HK will use 40,000 photomultiplier tubes (PMTs), thus reaching the same 40\,\% photocoverage as SK, and benefit from newly designed high-efficiency PMTs described below.

Construction is expected to take eight years, with start of operations planned for 2026. The option to add a second detector soon afterwards is actively being explored.

\subsection{Second Tank in Korea}
While the second detector could be located in Japan at the same site as the first one, the alternative possibility of building the second tank in Korea was explored in a white paper published in November 2016~\cite{T2HKK}.
In addition to sensitivity improvements for the long-baseline experiment, the Korean candidate sites offer a higher overburden (and thus lower spallation backgrounds) than the Japanese HK site, which would increase sensitivity of low-energy rare event searches like solar or supernova relic neutrinos.

\subsection{Photodetector Development}
A new 50 cm PMT model, the Hamamatsu R12860-HQE, was developed for HK. It is based on Hamamatsu’s R3600 PMT used in SK, but includes a box-and-line dynode and several other improvements. As a result, this new model offers better timing resolution and twice the detection efficiency due to improvements to both quantum efficiency and collection efficiency. Work to reduce the dark noise rate and design new PMT covers for pressure resistance is currently ongoing.

In addition to this baseline design, R\&D on alternative photosensor options like hybrid photo-detectors, LAPPDs and multi-PMT modules is ongoing.

\section{Supernova Neutrinos}\label{ch:sn}

\subsection{Galactic Supernova}
For a galactic supernova at a fiducial distance of 10\,kpc, HK will detect $\mathcal{O}(10^5)$ neutrinos within about 10\,s.
This high event rate enables HK to resolve fast time variations of the event rate, which could give us information on properties of the progenitor (like its rotation) or on details of the supernova explosion mechanism like the roles of turbulence, convection and the standing accretion shock instability, SASI, on which there is significant disagreement between different computer simulations~\cite{Janka2016}.

\subsection{Supernovae in Neighbouring Galaxies}
Due to its large volume, HK would be sensitive to supernova bursts in nearby galaxies as well, observing $2000-3000$ events per detector from a SN1987a-like supernova in the Large Magellanic Cloud or $\mathcal{O}(10)$ events from a supernova in the Andromeda galaxy.

\begin{figure}[htb]
\hspace{-0.0pc}\begin{minipage}{16.5pc}
\includegraphics[width=16.5pc]{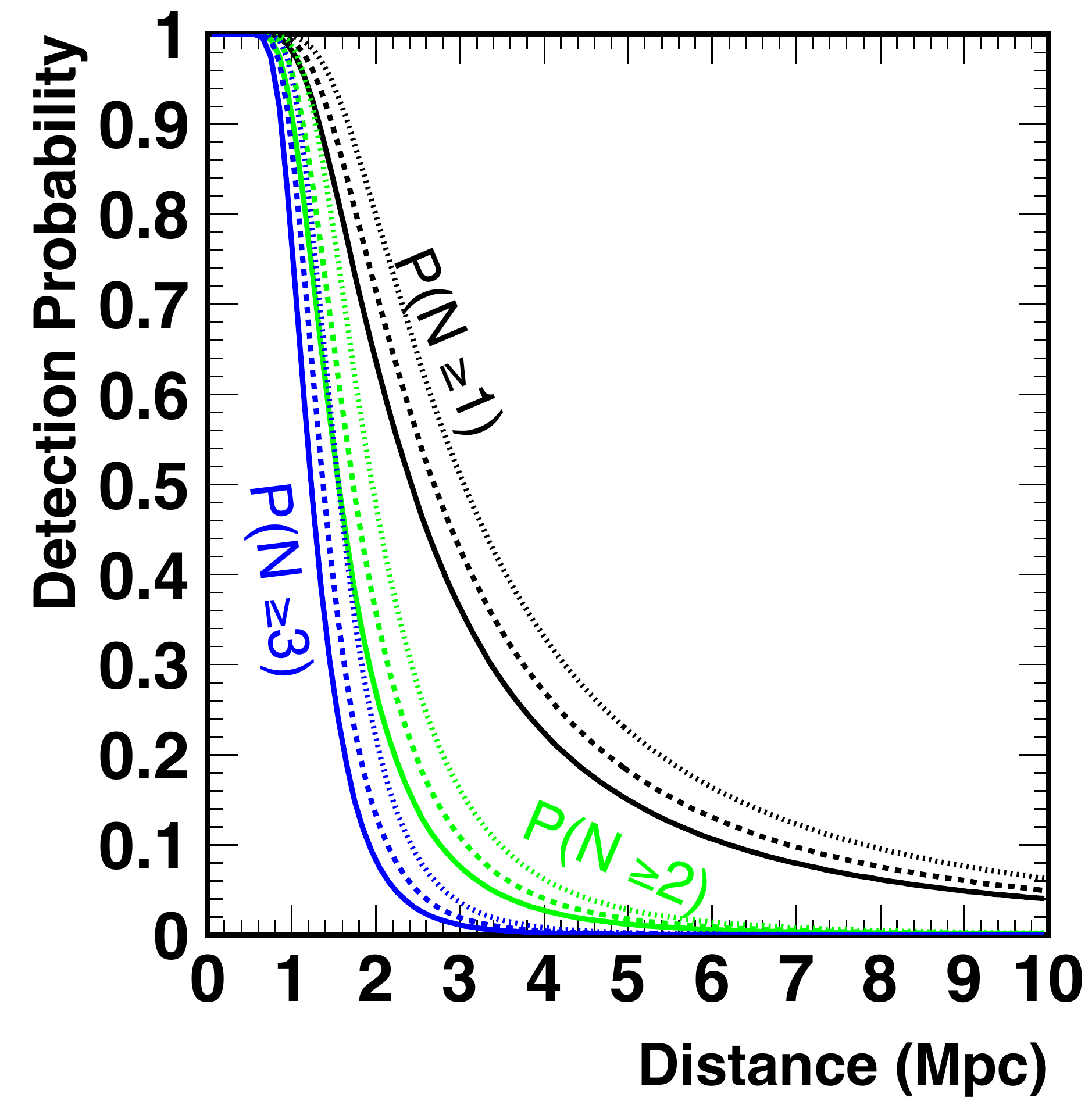}
\end{minipage}\hspace{2pc}
\begin{minipage}{16.5pc}
\includegraphics[width=16.5pc]{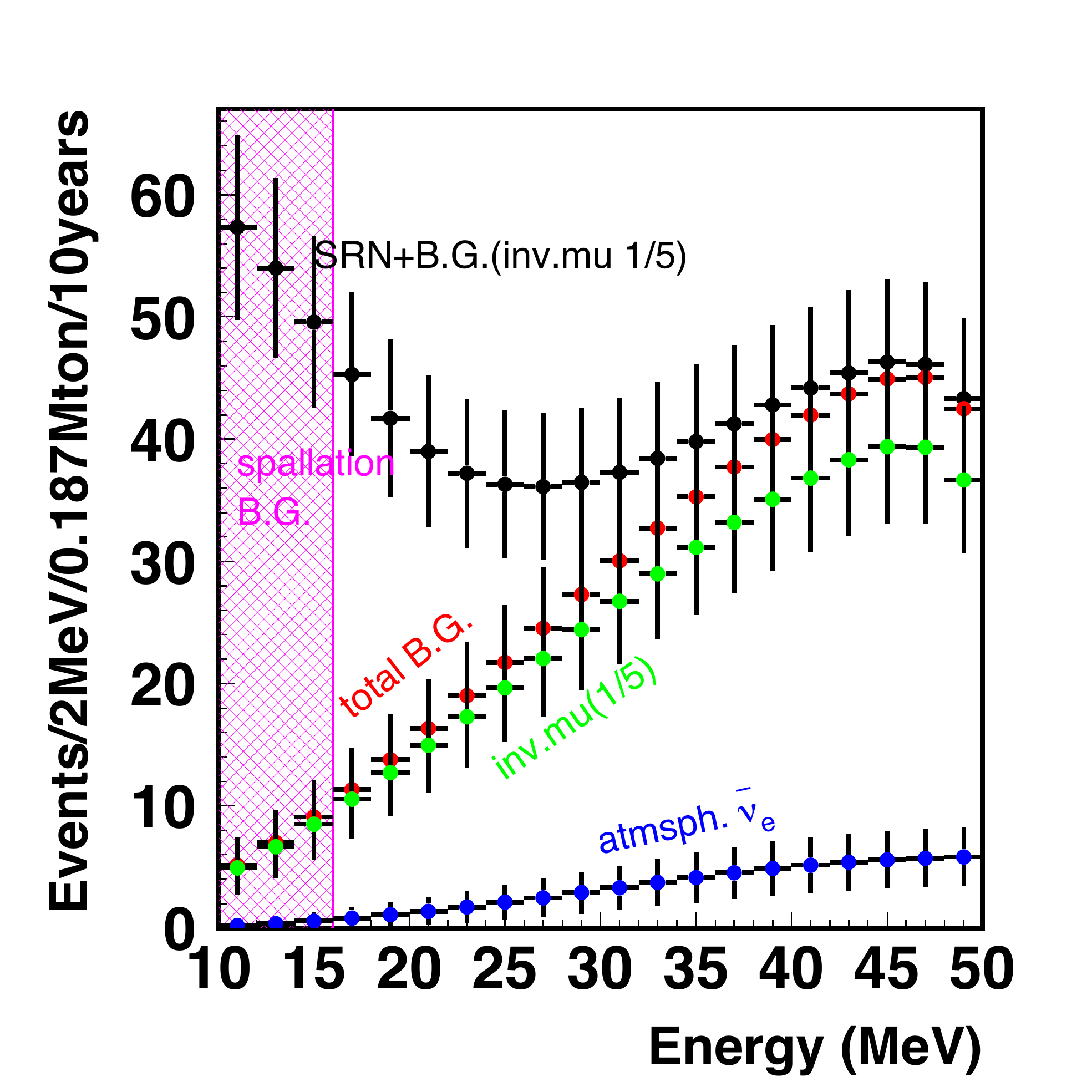}
\end{minipage} 
\caption{Left: Detection probability of supernova neutrinos versus distance at HK. Solid, dotted and dashed lines are for no oscillation, normal hierarchy and inverted hierarchy, respectively. Right: Expected spectrum of the SRN signal in HK after 10 years, including backgrounds.}
\label{fig:supernova}
\end{figure}

Using strict timing coincidence with an external trigger, like a gravitational wave signal in LIGO, VIRGO or the nearby KAGRA, HK could even be sensitive to single supernova neutrino events. For supernovae at up to 4\,Mpc distance, which are expected to happen every 3--4 years on average, HK has a 30\,\% or greater chance of detecting at least one event (see fig.~\ref{fig:supernova}).

\subsection{Supernova Relic Neutrinos}
The total neutrino flux from all remote supernovae in the history of the universe is known as supernova relic neutrinos (SRN). While these neutrinos cannot be traced back to a specific supernova, they deliver information on the average spectrum of supernova neutrinos and could enable a first measurement of the rate of failed (optically dark) supernovae, which are the origin of stellar mass black holes. The SRN measurement is therefore complementary to observations of nearby supernovae.

At HK, SRN are observable in an energy window of 16--30\,MeV (fig.~\ref{fig:supernova}), which is bounded by cosmic-ray induced spallation backgrounds at lower energies and invisible muon background from atmospheric neutrinos at higher energies. Within 10 (20) years, about 70 (140) SRN events are expected at HK, corresponding to an observation of SRN with 4.2 (5.7)\,$\sigma$ significance and allowing the first high-statistics measurement of the SRN spectrum.

\section{Solar Neutrinos}\label{ch:solar}
Solar {\it Hep} neutrinos from the reaction $^3 \textrm{He} + \textrm{p} \rightarrow\/ ^4 \textrm{He} + e^+ + \nu_e$ have never been detected due to the very low branching rate. Due to its high volume and the better energy resolution of the high-efficiency PMTs, which reduces backgrounds from $^8$B neutrinos, HK is expected to make the first detection of these neutrinos, which would be an excellent test of the standard solar model.

The unprecedented statistics of HK’s solar neutrino measurements enable us to look for variations of the solar neutrino flux on shorter time scales. 

In addition, solar neutrinos offer sensitive probes of the three neutrino mixing paradigm which are described in more detail in the remainder of this chapter.

\subsection{Day-Night Asymmetry}
At night, solar neutrinos travel through Earth to reach the detector and their oscillation probability is influenced by the matter potential they traverse. This partially regenerates the electron flavour content of the solar neutrino flux, which increases the observed event rate.

A simple way to define this asymmetry is $A_\textrm{DN} = 2\cdot (\Phi_\textrm{day} - \Phi_\textrm{night}) / (\Phi_\textrm{day} + \Phi_\textrm{night})$, which SK has measured to be $A_\textrm{DN} = (- 3.3 \pm 1.0\textrm{\,(stat.)\,} \pm 0.5\textrm{\,(syst.)\,})\,\% $, a $3\,\sigma$ evidence that is compatible with theoretical expectations based on the solar best-fit oscillation parameters~\cite{SK4solar}. There is, however, a $2\,\sigma$ tension between the values of $\Delta m^2_{12}$ determined by solar neutrino experiments on one side and KamLAND’s reactor antineutrino measurement on the other side. Assuming the KamLAND value, the theoretically predicted day-night asymmetry is $A^\textrm{KamLAND}_\textrm{DN} = - 1.7\,\%$.

Assuming the solar best fit values, HK would be able to detect the day-night asymmetry at the $5\,\sigma$ level within 3 years. Depending on the level of statistical uncertainty achieved, it would be able to distinguish between the solar and KamLAND values of $\Delta m^2_{12}$ at a level of 4--5\,$\sigma$ within 10 years. 

\subsection{Spectrum Upturn}
At energies below about 1\,MeV, the survival probability of solar neutrinos is determined by their vacuum oscillations on the way to Earth. 

At energies above about 10\,MeV, solar neutrinos undergo a flavour transition (MSW effect) due to the smoothly varying density of the matter they are traversing. They leave the sun in a mass eigenstate and do not undergo vacuum oscillations on their way to Earth.

In the transition region between 1 and 10\,MeV, the energy dependence of the survival probability is predicted to show a smooth upturn in standard three neutrino mixing, while new physics like non-standard interactions or sterile neutrinos could change the shape of this curve significantly. While current experiments are not sensitive to the shape of this curve~\cite{SK4solar}, HK would be able to detect this upturn at a 3--5\,$\sigma$ sensitivity level within 10 years.

\section{Indirect Dark Matter Searches}\label{ch:dm}

\begin{figure}[htb]
\hspace{-0.0pc}\begin{minipage}{17.5pc}
\includegraphics[width=17.5pc]{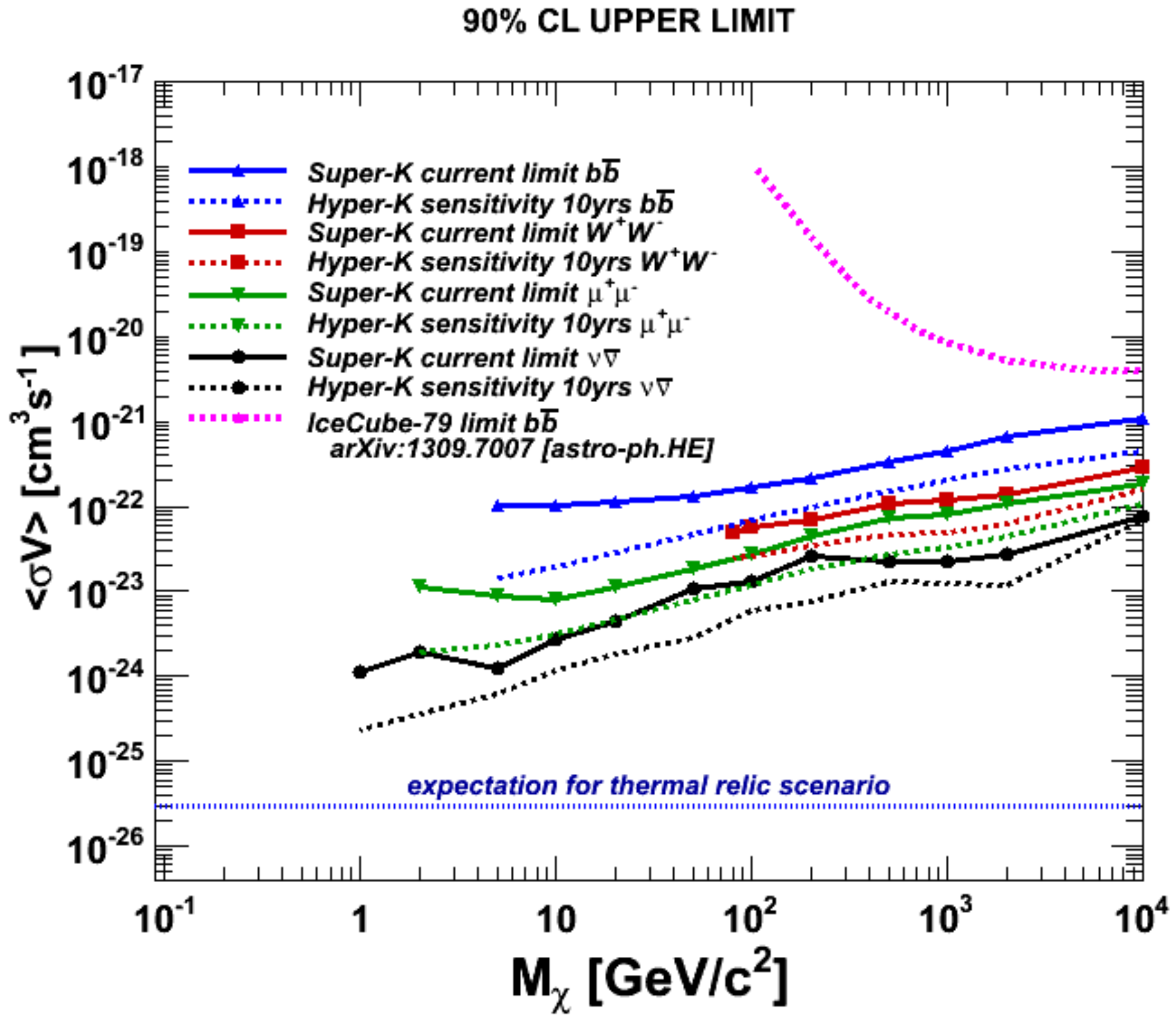}
\end{minipage}\hspace{1pc}
\begin{minipage}{17.5pc}
\includegraphics[width=17.5pc]{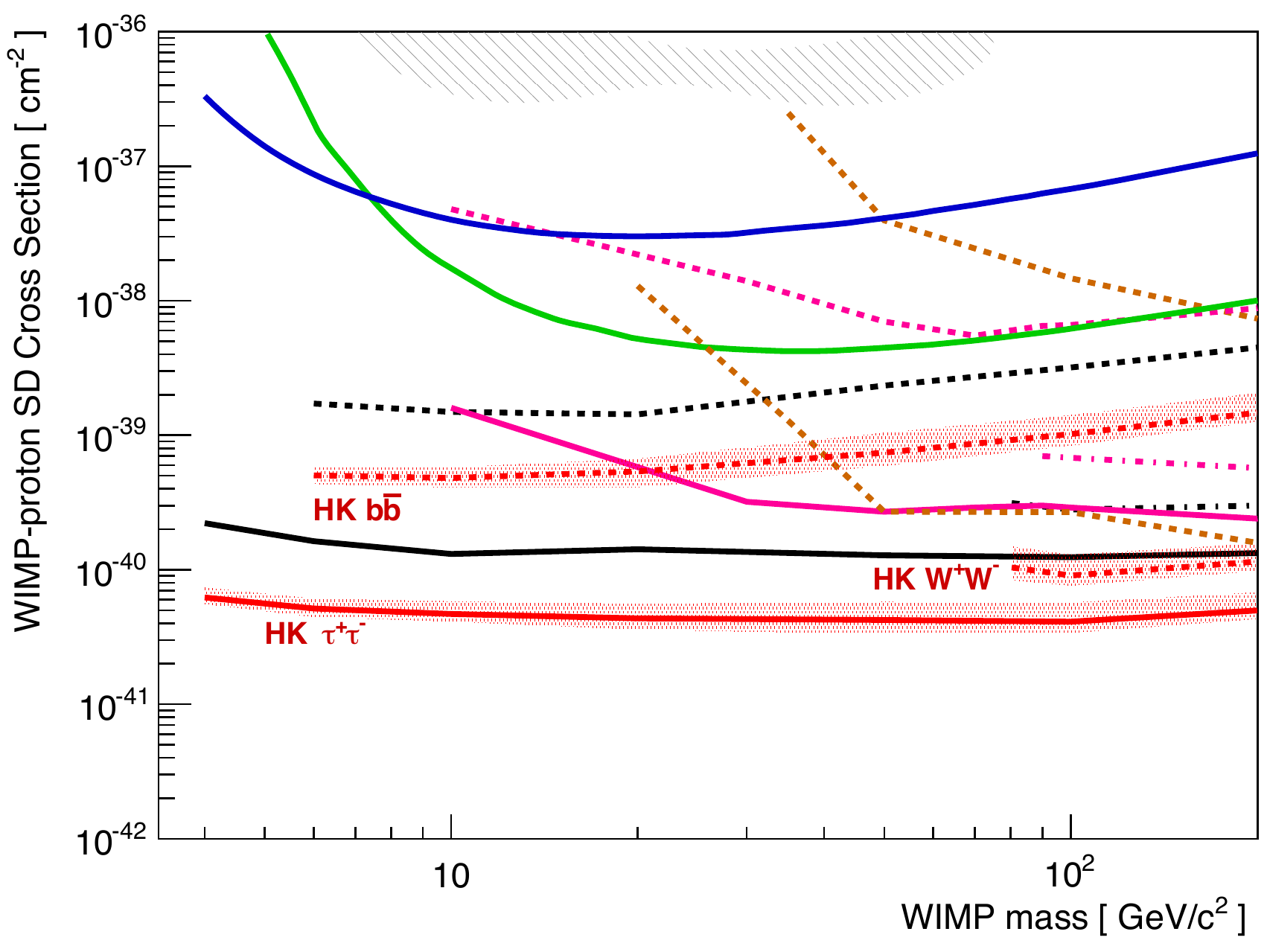}
\end{minipage} 
\caption{Left: Expected HK 90\,\% C.L. upper limits on the WIMP velocity averaged annihilation cross section for several annihilation modes at the galactic centre. Right: Expected HK 90\,\% C.L. upper limits on the spin-dependent WIMP-nucleon scattering cross section for neutrinos coming from the direction of the Sun. Black lines show current SK limits while other colours show limits from direct detection experiments.} 
\label{fig:dm}
\end{figure}

In many theoretical models, dark matter particles can co-annihilate, producing either neutrinos or other standard model particles that go on to decay into final states containing neutrinos.
Therefore, HK can perform indirect dark matter searches by looking for an excess flux of neutrinos from regions with a high dark matter density.
Furthermore, if the WIMP capture rate in the Sun is equal to the annihilation rate, limits on the WIMP-induced neutrino flux can be converted into limits on the WIMP-nucleon scattering cross section, allowing a comparison to sensitivities of direct detection experiments.
Figure~\ref{fig:dm} shows expected HK sensitivities for dark matter annihilation at the galactic centre and the Sun.
It shows that HK is particularly sensitive to lower WIMP masses, making it complementary to other dark matter detection experiments.

\begin{thebibliography}{9}
\bibitem{PublicDR}
  K.~Abe {\it et al.} [Hyper-Kamiokande Proto-Collaboration],
  \emph{Hyper-Kamiokande Design Report},
  KEK-Preprint-2016-21, ICRR-Report-701-2016-1.

\bibitem{T2HKK}
  K.~Abe {\it et al.} [Hyper-Kamiokande Proto-Collaboration],
  \emph{Physics Potentials with the Second Hyper-Kamiokande Detector in Korea},
  {\tt arXiv:1611.06118}.

\bibitem{Janka2016}
  H.-Th.~Janka, T.~Melson and A.~Summa,
  \emph{Physics of Core-Collapse Supernovae in Three Dimensions: a Sneak Preview},
  \emph{Annu.~Rev.~Nucl.~Part.~Sci.}~\textbf{66} (2016) p.\,341--375
  [{\tt arXiv:1602.05576}].

\bibitem{SK4solar}
  K.~Abe {\it et al.} [Super-Kamiokande Collaboration],
  \emph{Solar Neutrino Measurements in Super-Kamiokande-IV},
  \emph{Phys.~Rev.~D}~{\bf 94} (2016) no.\,5, 052010
  [{\tt arXiv:1606.07538}].

\end{thebibliography}
\end{document}